\documentclass[overfull,preprint]{epl}
\usepackage{psfrag}
\usepackage{array}
\usepackage{subfigure}
\usepackage{color} 

\title{DMRG evaluation of the Kubo formula -- \\%
{\large Conductance of strongly interacting quantum systems}}

\author{D. Bohr\inst{1,2} \and P. Schmitteckert\inst{2} \and P. W\"{o}lfle\inst{2}}
\institute{
  \inst{1} MIC -- Department of Micro and Nanotechnology \\
  Technical University of Denmark -- Building 345E, DK-2800 Kgs.~Lyngby, Denmark\\
  \inst{2} TKM -- Institut f\"{u}r Theorie der Kondensierten Materie\\
  Universit\"{a}t Karlsruhe, Wolfgang-Gaede-Str.~1, D-76128 Karlsruhe, Germany }


\begin{document}

\maketitle

\title{DMRG evaluation of Kubo formula}

\begin{abstract}
In this paper we present a novel approach combining linear response
theory (Kubo) for the conductance and the Density Matrix
Renormalization Group (DMRG). The system considered is
one-dimensional and consists of non-interacting tight binding leads
coupled to an interacting nanostructure via weak links. Electrons
are treated as spinless fermions and two different correlation
functions are used to evaluate the conductance.

Exact diagonalization calculations in the non-interacting limit
serve as a benchmark for our combined Kubo and DMRG approach in this
limit. Including both weak and strong interaction we present DMRG
results for an extended nanostructure consisting of seven sites. For
the strongly interacting structure a simple explanation of the
position of the resonances is given in terms of hard-core particles
moving freely on a lattice of reduced size.
\end{abstract}

\section{Introduction}
During the past decade improved experimental techniques have made
production of and measurements on one-dimensional systems possible
\cite{Sohn_Kouwenhoven_Schon:1997}, and hence led to an increasing
theoretical interest in these systems. Since its formulation in 1992
\cite{White:1992-1993} the Density Matrix Renormalization Group
method (DMRG) has been established as a very powerful, quasi-exact
method for numerical calculations of properties of (quasi)
one-dimensional systems.

In this paper we present a new approach for calculating linear
response conductance for one-dimensional interacting nanostructures
coupled to non-interacting tight binding leads. The method combines
Kubo expressions for the conductance with numerical DMRG
calculations and is valid for arbitrary interaction strength.
It facilitates  a unified description of strong and weak
interactions and provides  conductance directly form a transport calculation,
without relying on relations between equilibrium and transport properties.

We employ current-density and
current-current correlation functions to calculate the conductance
and in the non-interacting case compare to exact diagonalization
calculations.

In the strongly interacting limit a simple interpretation of the
position of the resonances is given in terms of freely moving hard-core
particles on a reduced size lattice \cite{Gomez_Santos:1993}, and
quantitative comparison with numerical DMRG results shows good
agreement.

\section{Model}
\begin{figure}[tb]
\onefigure[angle=90, width=0.8\textwidth]{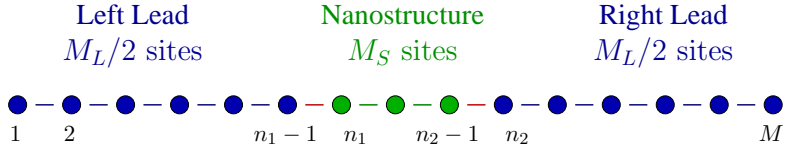}
\caption{One-dimensional interacting nanostructure with $M_S$ sites,
coupled to non-interacting tight binding leads. The total system
size is denoted $M$, the number of lead sites is $M_L$. The interdot
and interlead hopping elements are $t_{\textrm{Dot}}$ and $t$
respectively, while the contact between the nanostructure and leads
are via $t_L$ and $t_R$.} \label{Model_figure}
\end{figure}

We are interested in studying the effect of correlations on
transport within a microscopic model of an interacting
one-dimensional nanostructure coupled to two non-interacting tight
binding leads, as shown in Fig.~\ref{Model_figure}. Electrons are
treated as spinless and only nearest neighbor interaction is
considered. The corresponding Hamiltonian is
\begin{eqnarray}
  \hat{H}_0 &=& \hat{H}_{\mathrm{NS}} \,+\, \hat{H}_{\mathrm{L}} \,+\,\hat{H}_{\mathrm{C}},\label{Equilibrium Hamiltonian}\\
  \hat{H}_{\mathrm{NS}} &=& \sum_{j=n_1}^{n_2-1}U_g c_j^\dagger c_j^{\phantom\dagger} +\sum_{j=n_1+1}^{n_2-1}\big(-t_{\rm Dot}(c_j^\dagger c_{j-1}^{\phantom\dagger} + c_{j-1}^\dagger c_j^{\phantom\dagger})+ V c_j^\dagger c_j^{\phantom\dagger} c_{j-1}^\dagger c_{j-1}^{\phantom\dagger}\big),\\
  \hat{H}_{\mathrm{L}} &=& -t\sum_{i=2}^{n_1-1}(c_i^\dagger c_{i-1}^{\phantom\dagger} + c_{i-1}^\dagger c_i^{\phantom\dagger})-t\sum_{i=n_2+1}^{M}(c_i^\dagger c_{i-1}^{\phantom\dagger} + c_{i-1}^\dagger c_i^{\phantom\dagger}),\\
  \hat{H}_{\mathrm{C}} &=& -t_L(c_{n_1}^\dagger c_{n_1-1}^{\phantom\dagger}+c_{n_1-1}^\dagger c_{n_1}^{\phantom\dagger})-t_R(c_{n_2}^\dagger c_{n_2-1}^{\phantom\dagger}+c_{n_2-1}^\dagger c_{n_2}^{\phantom\dagger})\nonumber\\
  && + \gamma_V V (c_{n_1}^\dagger c_{n_1}^{\phantom\dagger}c_{n_1-1}^\dagger c_{n_1-1}^{\phantom\dagger}+c_{n_2}^\dagger c_{n_2}^{\phantom\dagger}c_{n_2-1}^\dagger c_{n_2-1}^{\phantom\dagger}).
\end{eqnarray}
The parameter $\gamma_V$ controls the smoothing of the interaction
on the dot over the contact links as discussed in
\cite{Molina_Schmitteckert_Weinmann_Jalabert_Ingold_Picard:2004},
and $U_g$ is a gate voltage on the structure. In this work we set $t = t_{\rm Dot}=1$.

\section{Kubo Expressions}
Using linear response in applied source-drain voltage, $V_{SD}(t)$,
the current is given by
\begin{eqnarray}
  \hat{H} &=& \hat{H}_0+\delta \hat{H},\\
  \big<\tilde J_n(t)\big> &=& \bar J-i\int_{-\infty}^t dt' \big<\psi_0\big|[\tilde J_n(t),\delta \tilde H(t')]
  \big|\psi_0\big>,\\
  \tilde{J}_{n}(t) &=& -it_{n} \big[ \tilde{c}_{n}^\dagger(t) \tilde{c}_{{n}-1}^{\phantom\dagger}(t)-\tilde{c}_{{n}-1}^\dagger(t)
  \tilde{c}_{n}^{\phantom\dagger}(t)\big],
\end{eqnarray}
where $\hat{H}_0$ is the Hamiltonian in eq.~\ref{Equilibrium
Hamiltonian}, the applied voltage perturbation is $\delta \hat
H(t)=V_{SD}(t) \hat{N}$, $\tilde A(t)=e^{i \hat{H}_0 t} \hat{A}
e^{-i \hat{H}_0 t}$ denotes the interaction picture time evolution
of the operator $\hat{A}$, and $\big|\psi_0\big>$ denotes the ground
state. Note that in this approach $\tilde A(t)$ contains all
correlations of the unbiased structure, since we apply it to the
quasi-exact ground state given by the DMRG procedure. The number
operator is taken as a symmetric combination of the left and right
lead operators, $\hat N=\frac 12(\hat N_L-\hat N_R)$, and $\bar J$
is the equilibrium current included for completeness and henceforth
neglected in all numerical calculations.

The Kubo conductance in the DC limit, $g\equiv
\frac{e^2}{h}\big<\tilde J\big>/V_{SD}$, can be expressed in
terms of two different correlators,
\begin{eqnarray}
  g_{J_jN} &=& -\frac{e^2}{h}\big<\psi_0\big|\hat{J}_{n_j}
  \frac{4\pi i\eta}{(\hat H_0-E_0)^2+\eta^2}\hat N\big|\psi_0\big>,\label{g_JV definition}\\
  g_{JJ} &=& \frac{e^2}{h}\big<\psi_0\big|\hat{J}_{n_1}\frac{8\pi \eta(\hat H_0-E_0)}{\big[(\hat H_0-E_0)^2+\eta^2\big]^2}\hat{
  J}_{n_2}\big|\psi_0\big>,\label{g_JJ definition}
\end{eqnarray}
where the positions $n_1$ and $n_2$ are defined in
Fig.~\ref{Model_figure}. Analogous Kubo expressions were used by
Louis and Gros in \cite{Louis_Gros:2003}, where Quantum Monte Carlo
calculations for the density-density correlator were performed.

\section{Damped Boundary Conditions}
To improve the finite size scaling and to facilitate the use of
sufficiently short leads we use exponentially damped boundary
conditions, decreasing the hopping elements towards the end of the
leads exponentially as shown for the right lead in eq.~\ref{Damped
Boundary Conditions},\footnote{Modified BC's in connection with DMRG
were introduced by Vekic and White in \cite{White_Vekic:1993-1996}
using soft boundary conditions to reduce finite size effects. Note
that exponential damping corresponds to the hopping Hamiltonian in
the Numerical Renormalization Group, which models the logarithmic
discretization.}
\begin{eqnarray}
    [-t, \cdots, -t,\underbrace{-t,-t,\cdots,-t}_{M_D}] &\rightarrow&
    [-t, \cdots, -t,\underbrace{-td,-td^{2},\cdots,-td^{M_D-1},
    -td^{M_D}}_{M_D}]\,,\label{Damped Boundary Conditions}
\end{eqnarray}
where $d<1$. The improvement of the finite size scaling relies on
two properties of the DBC's: (1) They allow for use of a smaller
$\eta$ and (2) serve as a particle bath for the nanostructure. The
first property is caused by the introduction of exponentially small
energyscales in the system thus reducing the finite size level
splitting at the Fermi energy at half filling. The second property
can be understood from the fact that the energy cost of adding or
removing a particle from the damped region is of the order of the
exponentially small hopping element.\footnote{In principle
properties (1) and (2) of the DBC's could be obtained by using
longer non-damped leads. However these leads would have to be
\emph{exponentially long} making such a direct approach impossible.}

The DBC's introduce two more parameters in the model, the number of
damped bonds $M_D$ and the damping factor $d$, and these must take
values such that physical quantities do not depend sensitively on
the particular choice.

\section{Numerical Calculations}
Before actual numerical calculations can be performed the parameters
of the model, $M_D$, $d$, and $\eta$, must be determined. This is
done using exact diagonalization calculations for the
non-interacting systems, specifically the resonant value at
$U_g=0$.\footnote{Considering structures consisting of an odd number
of sites has the advantage that the central resonance (by symmetry)
remains at $U_g=0$ for half filled leads. Due to the bath property
of the DBC's it is safe to assume that half filling is maintained in
the parts of the leads that are close to the nanostructure. In contrast
the strongly damped regions act like particle baths and therefore
cannot maintain half filling for non-zero external potential.} For
fixed $M_D$ we do indeed find a range of $d$ values that produce
essentially identical physical results, indicating the range of
validity of the DBC's. Additionally we find that the actual value of
$M_D$ is not significant (for reasonably large values) as long as
the corresponding value of $d$ is tuned such that the damping at the
edge reaches values of the same order of magnitude. The leads used are
sufficiently long to keep the damped region separated from the
nanostructure, thus allowing Friedel oscillations at the structure
edge to decay before reaching the damped region.

The magnitude of the parameter $\eta$ is bounded by physical
arguments; from below by the fact that it should be larger than the
finite size level splitting to allow transport, and from above by
the broadening of physical results by any finite $\eta$, and should
thus be much smaller than the width of the resonances we wish to
resolve. It is important to note, that $\eta$ is an inherent property of any
transport calculation and can only be avoided if one finds a way
to obtain transport properties from equilibrium properties.

The conductances in eqs.~\ref{g_JV definition} and \ref{g_JJ
definition} are given in terms of ground state correlators and hence
DMRG is directly applicable. To evaluate the correlators we use the
correction vector DMRG
\cite{Kuhner_White:1999,Jeckelmann:2002,Ramasesha:1990} in the zero
frequency limit. Calculating, e.g., the correlator in eq.~\ref{g_JJ
definition} is done by formulating the linear problems,
\begin{equation}
  \frac{1}{\hat H_0-E_0+i\eta}\hat J_{n_j}\big|\psi_0\big> \;=\;
  \big|\phi_j\big>\quad\Rightarrow\quad
  \hat J_{n_j}\big|\psi_0\big> \;=\; \big[\hat
  H_0-E_0+i\eta\big]\big|\phi_j\big>,\label{Correction vector
  definition}
\end{equation}
which can be solved for $\big|\phi_j\big>$ by a linear solver.
Having solved for the correction vector $\big|\phi_j\big>$ the
conductance is found as the vector overlap,
\begin{eqnarray}
  \big|\phi_j\big> &=& \big|\phi_j^R\big> + i \big|\phi_j^I\big>,\\
  g_{JJ} &=& -\frac{8\pi e^2}{h}\big<\phi_1^I\big|\phi_2^R\big>.
\end{eqnarray}
In our DMRG calculations we target apart from the ground state also
the real and imaginary parts of the two correction vectors,
$\big|\phi_1\big>$ and $\big|\phi_2\big>$, as well as the states
$\hat N\big|\psi_0\big>$ and $\hat J_{n_{1,2}}\big|\psi_0\big>$ to
ensure that the DMRG basis is suitable for describing the
conductance accurately \cite{Kuhner_White:1999,Jeckelmann:2002}.

It should be mentioned that the damped boundary conditions make the
convergence rate in numerical calculations much slower. In addition
any finite external gate voltage, $U_g$, changes the particle number
in the structure and the excess particles come from the bath
property of the DBC's. We therefore face the problem that the
damping should be sufficiently strong to provide a reasonable
particle bath but at the same time a strong damping decreases the
coupling of the highly damped region to the rest of the system. To
remedy the slow convergence in the DMRG calculations we turn on the
damping in steps and perform several finite system DMRG sweeps for
each such damping step. In other words, we perform a complete finite
lattice calculation employing typically 11 sweeps and then initiate
the scaling sweeps. This allows DMRG to gradually optimize the basis
to include the damping in the leads and provides a more gradual
decoupling of the damped regions from the rest of the system, thus
improving the convergence rate at the cost of more DMRG iterations.

\begin{figure}[tbp]
\begin{center}
    \newlength{\graphiclength}
    \setlength{\graphiclength}{0.475\textwidth}
    \subfigure[Single resonant level, $M_S=1$ and $M=102$. $f$'s denote
    exact diagonalization results, $g$'s denote DMRG
    results, and $L$ denotes the exact Lorentzian  in the
    infinite lead limit. The inset shows an enlargement of the resonance
    peak.]{\label{MS=1,V=0.0}
    \includegraphics[width=\graphiclength]{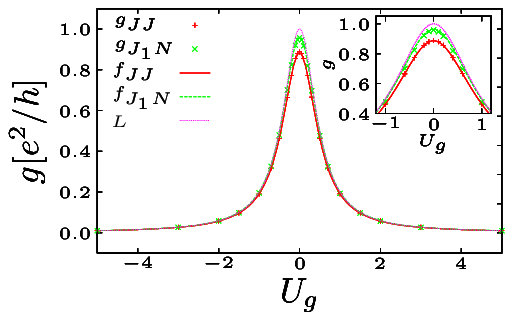}}
    \hspace{0.02\textwidth}\subfigure[Seven site nanostructure, $M_S=7$ and $M=150$,
    in the non interacting limit, $V=0.0$. Exact diagonalization calculation.]
    {\label{MS=7,V=0.0}
    \includegraphics[width=\graphiclength]{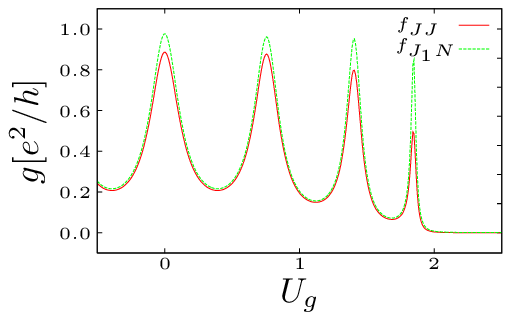}}\\
    \subfigure[Seven site nanostructure, $M_S=7$ and $M=150$, in the
    Luttinger Liquid regime, $V=1.0$.]{\label{MS=7,V=1.0}
    \includegraphics[width=\graphiclength]{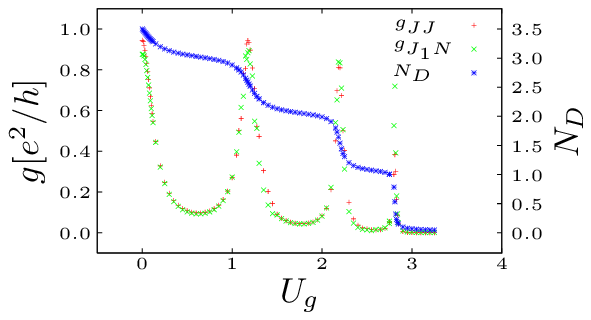}}
    \hspace{0.02\textwidth}
    \subfigure[Seven site nanostructure, $M_S=7$ and $M=150$, in the charge density wave
    regime, $V=5.0$.]
    {\label{MS=7,V=5.0}
    \includegraphics[width=\graphiclength]{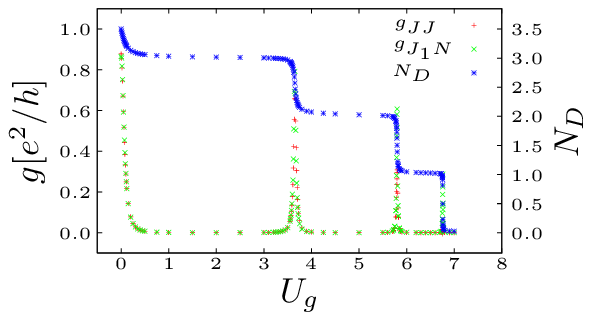}}
    \caption{Conductance, $g$, and number of particles on the dot, $N_D$,
    versus external potential $U_g$ for a single resonant level and for
    an extended nanostructure consisting of seven sites. $g_{JJ}$
    denotes the current-current correlator, and $g_{J_jN}$ denotes
    the current-density correlator. The left/right
    contact hopping elements are $t_{L/R}=0.5$ and the parameters of the
    DBC's are $M_D=30$ and $d=0.8$. For the calculations above we use $\eta=1/M$.
    For the interacting spectra notice the offset of resonance positions
    of the order $V$ as compared to the noninteracting case. The interaction
    on the nanostructure is smoothed over the contacts with $\gamma_V=0.5$.}\label{AllFigures}
\end{center}
\end{figure}

Nevertheless the resolvent equations, eq.~\ref{Correction vector
definition}, are still ill-conditioned and standard solvers like the
Conjugate Gradient Method do not converge. We use instead a
preconditioned Davidson type solver similar to Ramasesha
\cite{Ramasesha_Pati_Krishnamurthy_Shuai_Bredas:1989} modified
with a Gauss-Seidel enhanced block diagonal preconditioner.
The DMRG calculations presented in Fig.~\ref{AllFigures} were done
using up to $m=1200$ states. In our DMRG implementation we do not
fix the number of states per block to be $m$ but rather fix the
dimension of the target space to be at least $m^2$. In the
calculations presented this corresponds to an increase of block states
of typically $15\%-30\%$.

\section{Results}
Here we present DMRG and (in the non-interacting limit) exact
diagonalization calculations for a single resonant level,
Fig.~\ref{MS=1,V=0.0}, and a nanostructure consisting of seven sites
coupled symmetrically to two non-interacting leads. For the extended
structure we present results in the non interacting limit,
Fig.~\ref{MS=7,V=0.0}, and for weak and strong interaction,
Fig.~\ref{MS=7,V=1.0} and \ref{MS=7,V=5.0}.

The spinless single resonant level is generically non-interacting
and serves as a testing ground for the approach. The exact result
for the conductance in the symmetrically coupled case can be shown
to be a Lorentzian of full width $4t'^2$ at half maximum, where
$t'=t_L=t_R$. In Fig.~\ref{MS=1,V=0.0} we show exact diagonalization
and DMRG calculations for the single resonant level and the two sets
are virtually indistinguishable. This verifies that the truncation
error introduced by the DMRG is negligible. Furthermore we have
plotted the exact Lorentzian result, and the agreement between the
three curves is very good, demonstrating the accuracy of our
combined Kubo and DMRG approach.

There is a systematic difference between the current-current and the
current-density correlators, specifically close to resonances the
current-density correlator generally gives better results. This is
due to the additional energy dependent broadening given by $\hat
H_0-E_0$ in the current-current correlator.
The opposite is true in the tails where the
current-current correlator is more reliable since it is less
sensitive to changes of the particle number.

In numerical calculations the parameter $\eta$ is always finite
making the expected form of the conductance peaks that of an area
normalized Lorentzian ($L_A$) of half-width $\eta$ convoluted with
the ``bare'' physical result. Assuming as a first approximation that
the latter is a height normalized Lorentzian ($L_H$) of width
$\Gamma$, the expression for the expected numerical results is of
the general form
\begin{eqnarray}
(L_A*L_H)(x)
    = \frac{\Gamma}{2}\frac{\eta + \Gamma/2}{(x-x_0)^2+(\eta + \Gamma/2)^2} 
.
\end{eqnarray}
To leading order in $\eta$ the conductance at the resonant level is
then given by $g_{\textrm{res}}\approx 1 - {2\eta}/{\Gamma}$, which
demonstrates that one needs small $\eta$ to reach the unitary limit,
$g_{\textrm{res}}=1$. However $\eta$ is known from the input and
$\Gamma$ can be extracted from the results. Thus a conductance value
on resonance of $1-2\eta/\Gamma$ is explained entirely by the
broadening by the finite leads and therefore suggests that infinite
leads in this case would yield the unitary limit. Our calculations
indicate, that the peak width is only slightly affected by the
interaction on the nanostructure, as long as the nanostructure
remains in the Luttinger liquid regime. However, once the structure
is driven into the charge density wave regime the peak width
decreases rapidly. A more detailed study of the resonance shapes is
considered future work.

The position of the resonances can be described by the addition
spectrum,
\begin{eqnarray}
   U_g^{N_D-1\to N_D} &=& E_0^{N_D-1}-E_0^{N_D},\label{Addition Spectrum U_g}
\end{eqnarray}
where $E_0^{N_D}$ is the energy of the isolated nanostructure occupied
by $N_D$ particles. In the large interaction limit the kinetic energy
of the particles can be approximated by freely moving fermions on an
effective lattice of size $M^*_S=M_S-N_D$. In this approximation one
describes the interacting fermions by effective hard-core particles
of the size of the interaction range, compare
\cite{Gomez_Santos:1993,Schmitteckert_Werner:2004}. Thus
eq.~\ref{Addition Spectrum U_g} can be expressed as
\begin{eqnarray}
  U_g^{N_D-1\to N_D} &=& V+2t\Big(\sum_{n=1}^{N_D}\cos(\frac{\pi n}{M_S-N_D})-\sum_{n=1}^{N_D-1}\cos(\frac{\pi
  n}{M_S-(N_D-1)})\Big),\label{Predicted Resonance Positions}
\end{eqnarray}
where $N_D$ should be small enough that the nanostructure is
still in a delocalized state.

In an effective charging model the additional splitting of the
levels due to the interaction is linear in the charging interaction
$V$. By contrast, in our microscopic model the interaction leads to an
overall \emph{offset} for the non-central peaks, while their mutual
\emph{splitting} is governed by the kinetic energy, $\sim t$.

In tab.~\ref{Peak Position Table} we show a comparison of resonance
positions as predicted by the reduced lattice (RL) model in
eq.~\ref{Predicted Resonance Positions}, as predicted by exact
diagonalization (ED) of the isolated nanostructure, and resonances
found in our DMRG calculations for interaction strengths $V=5$,
$20$, $30$. The position of the outermost resonance from $0\to 1$
particle fits fairly well for both predictions, while the next ones
deviate somewhat. The RL prediction for the transition $2\to 3$ is
not expected to be accurate since $N_D=3$ is a localized charge
density wave like state. All exact diagonalization predictions are
correct to lowest order in $t/V$ as expected.
\begin{table}
\caption{Table of peak positions for the $M_S=7$ site structure with
interaction $V=5,20,30$, as predicted by the reduced lattice (RL)
model, by exact diagonalization (ED) of the isolated nanostructure,
and as found from the conductance peaks in our DMRG calculations.
The RL prediction for $N_D=3$ is not expected to be accurate
since the nanostructure is in a localized charge density wave like
state.
Except for the RL prediction for $V=5$, $N_D=3$, all
predictions are correct to linear order in $t/V$.}
\label{Peak Position Table}
\begin{center}
\setlength{\extrarowheight}{0.5mm}
\begin{largetabular}{||l|c|c|c|c|c|c|c|c|c||}
\hline $V$ & \multicolumn{3}{c|}{5}  & \multicolumn{3}{c|}{20} & \multicolumn{3}{c||}{30}\\
\hline $N_D$ & 1 & 2 & 3  & 1 & 2 & 3 & 1 & 2 & 3\\
\hline $U_g^{N_D-1\to N_D}$ RL & 6.73 & 5.50 & 2.76   & 21.73  & 20.50 &  17.76  & 31.73  &  30.50  & 27.76 \\
\hline $U_g^{N_D-1\to N_D}$ ED & 6.77 & 5.88 & 3.85   & 21.75  & 20.63 & 18.03   & 31.74  & 30.59 &  27.94 \\
\hline $U_g^{N_D-1\to N_D}$ DMRG & 6.76 & 5.79 & 3.66 &  21.74 & 20.59 & 17.97  &  31.74 & 30.60 & 27.95\\
\hline
\end{largetabular}
\end{center}
\end{table}

\section{Conclusion}
In this work we have presented a new approach for linear conductance
calculations of interacting one-dimensional nanostructures,
combining linear response for conductance and DMRG. We have
benchmarked this new approach against exact diagonalization
calculations in the non-interacting case and found excellent
agreement, which serves as a real test for the real space DMRG. For
the resonant level we also compared our results to the exact
Lorentzian result, and found excellent agreement.

For the interacting case we have presented conductance curves for a
seven site nanostructure in both the Luttinger Liquid ($V=1$) and
the charge density wave ($V=5$) regimes, thus demonstrating the
versatility of our approach. We find the largest conductance when
the particle number in the structure fluctuates, in agreement with
physical intuition.

In the large interaction limit we have shown that a simple picture
based on effective hard-core particles moving freely on a reduced
size lattice describes the position of the resonances quite well.
However, the peak width is strongly decreased by strong interaction.

We expect that further finetuning of the method and numerical
parameters will lead to significantly more precise results
facilitating calculations for more complicated structures and allow
to quantitatively describe resonance peaks for strongly interacting and
extended structures.

\acknowledgments This work was performed at TKM, Universit\"{a}t
Karlsruhe and we profited from many discussions with colleagues. In
particular we would like to thank Ferdinand Evers,  Gert-Ludwig
Ingold, G\"unter Schneider, and Ralph Werner for their help in
clarifying concepts. D.~B.~is grateful for the hospitality of TKM
during this work. The authors acknowledge the support from the DFG
through project B2.10 of the Center for Functional Nanostructures,
and from the Landesstiftung Baden-W\"{u}rttemberg under project B710.


\begin{thebibliography}{srt}

\bibitem{Sohn_Kouwenhoven_Schon:1997}
  \Book{Mesoscopic electron transport: Proceedings of the NATO Advanced Study Institute} \Editor{Sohn L. L., Kouwenhoven L. P. \and Sch\"{o}n G.}
   \Publ{Kluwer}
   \Year{1997}

\bibitem{White:1992-1993}
  \Name{White S. R.}
  \REVIEW{Phys. Rev. Lett.}{69}{1992}{2863},
  \REVIEW{Phys. Rev. B}{48}{1993}{10345}.

\bibitem{Molina_Schmitteckert_Weinmann_Jalabert_Ingold_Picard:2004}
  \Name{Molina R. A., Schmitteckert P., Weinmann D., Jalabert R. A., Ingold G.-L. \and Picard J.-L.}
  \REVIEW{Eur. Phys. Jour. B}{39}{2004}{107}.

\bibitem{Louis_Gros:2003}
    \Name{Louis K. \and Gros C.}
    \REVIEW{Phys. Rev. B}{68}{2003}{184424}.

\bibitem{White_Vekic:1993-1996}
  \Name{White S. R. \and Vekic M.}
  \REVIEW{Phys. Rev. Lett.}{71}{1993}{4283},
  \REVIEW{Phys. Rev. B}{53}{1996}{14552}.

\bibitem{Ramasesha:1990}
    \Name{Ramasesha S.}
    \REVIEW{J. Comp. Chem.}{11}{1990}{545}.

\bibitem{Kuhner_White:1999}
    \Name{K\"{u}hner T. D. \and White S. R.}
    \REVIEW{Phys. Rev. B}{60}{1999}{335}.

\bibitem{Jeckelmann:2002}
  \Name{Jeckelmann E.}
  \REVIEW{Phys. Rev. B}{66}{2002}{45114}.

\bibitem{Ramasesha_Pati_Krishnamurthy_Shuai_Bredas:1989}
    \Name{Ramasesha S., Pati S. K., Krishnamurthy H. R., Shuai Z. \and Br\'{e}das J. L.}
    \REVIEW{Synth. Metals}{85}{1997}{1019}.

\bibitem{Gomez_Santos:1993}
    \Name{G\'{o}mez-Santos G.}
    \REVIEW{Phys. Rev. Lett.}{70}{1993}{3780}.

\bibitem{Schmitteckert_Werner:2004}
    \Name{Schmitteckert P. \and Werner R.}
   \REVIEW{Phys. Rev. B}{69}{2004}{195115}.

\end{thebibliography}
\end{document}